\documentclass[11pt]{article}

\usepackage{amsmath}
\usepackage{amsfonts}
\usepackage{dsfont}

\usepackage{color}

\usepackage{amssymb}

\usepackage{hyperref}

\def\be{\begin{equation}}
\def\ee{\end{equation}}
\def\bea{\begin{eqnarray}}
\def\eea{\end{eqnarray}}

\newcommand{\sect}[1]{\setcounter{equation}{0}\section{#1}}
\newcommand{\subsect}[1]{\subsection{#1}}

\def\k{\omega}

\def\C2v{{\widetilde C}_\vt} 
\def\S2v{{\widetilde S}_\vt}

\def\vt{\vartheta} 
\def\R{R}

\def\LL{L}

\def\producto{\cdot}                 
\def\>#1{{\mathbf #1}}

\begin{document}

\ 

 \vskip2cm

\begin{center}

 {\Large{\bf {The $\kappa$-(A)dS quantum algebra in (3+1) dimensions}}}

\medskip 
\medskip 
\medskip

{\sc \'Angel Ballesteros$^1$,  Francisco J.~Herranz$^1$,  Fabio Musso$^{2}$ and Pedro Naranjo$^1$}

\medskip

{$^1$ Departamento de F\'\i sica,  Universidad de Burgos,
E-09001 Burgos, Spain  \\
$^2$ Istituto Comprensivo ``Leonardo da Vinci",
Via della Grande Muraglia 37,
 I-0014 Rome, Italy \\
}
\medskip

\noindent
 E-mail: {\tt   angelb@ubu.es,    fjherranz@ubu.es,  fmusso@ubu.es, pnaranjo@ubu.es}

\end{center}

  \medskip 
\bigskip
\bigskip

\begin{abstract} 
\noindent The quantum duality principle is used to obtain explicitly the Poisson analogue of the  $\kappa$-(A)dS quantum algebra in (3+1)  dimensions as the corresponding Poisson--Lie structure on the dual solvable Lie group. The  construction is fully performed in a kinematical basis and deformed Casimir functions are also explicitly obtained. The cosmological constant $\Lambda$ is included as a Poisson--Lie group contraction parameter, and the limit $\Lambda\to 0$ leads to the well-known $\kappa$-Poincar\'e algebra in the bicrossproduct basis. A twisted version with Drinfel'd double structure of this $\kappa$-(A)dS deformation is sketched.
 \end{abstract}

\bigskip\bigskip\bigskip\bigskip


\bigskip

\noindent
KEYWORDS:  

\noindent anti-de Sitter, cosmological constant, quantum groups, Poisson--Lie groups, Lie bialgebras, quantum duality principle.


\section{Introduction}

Since Classical Gravity is essentially a theory describing the geometry of spacetime, it seems natural to consider that  a suitable definition of a ``quantum" spacetime geometry emerging at the Planck energy regime could be a reasonable feature of Quantum Gravity. Indeed, specific mathematical frameworks for such ``quantum geometry" have to be proposed. In particular,  ``quantum spacetime" is frequently introduced as a noncommutative algebra whose noncommutativity is governed by a parameter related to the Planck scale, thus leading to minimum length frameworks through generalized spacetime uncertainty relations (see, for instance,~\cite{Snyder, Yang, Maggiore, DFR, Garay} and references therein).

In this context, quantum groups~\cite{Drinfelda, CP, majid} provide a consistent approach to noncommutative spacetimes, since the latter are obtained as  noncommutative algebras that are covariant under the action of quantum kinematical groups. For instance, the well-known $\kappa$-Minkowski spacetime~\cite{kMinkowski, bicross2, kZakr, LukR} was obtained as a byproduct of the $\kappa$-Poincar\'e quantum algebra, which was introduced in~\cite{Lukierskia} (see also~\cite{Lukierskib, Giller, Lukierskic}) by making use of quantum group contraction techniques~\cite{CGSTe3, LBC}. One of the main features of the $\kappa$-Poincar\'e quantum algebra (which is the Hopf algebra dual to the quantum Poincar\'e group, and is defined as a deformation of the Poincar\'e algebra in terms of the dimensionful parameter $\kappa$) consists in its associated deformed second-order Casimir, which leads to a modified energy-momentum dispersion relation. From the phenomenological side, this type of deformed dispersion relations have been proposed as possible experimentally testable footprints of quantum gravity effects in very different contexts (see~\cite{AEN, Amelinodispersion1, Mattingly} and references therein).

Moreover, if the interplay between quantum spacetime and gravity at cosmological distances is to be modeled, then the curvature of spacetime cannot be neglected and models with non-vanishing cosmological constant have to be considered~\cite{Marciano, AMMR, FRW, iceCUBE}. Thus, the relevant kinematical groups (and spacetimes) would be the (anti-)de Sitter ones, hereafter (A)dS, and the construction of quantum (A)dS  groups should be faced. In (1+1) and (2+1) dimensions, the corresponding $\kappa$-deformations have been constructed~\cite{ck2, ck3} (see also~\cite{Zakr, tallin, LukiBorowiec} for classification approaches). In fact, it is worth stressing that the  $\kappa$-(A)dS deformation introduced in~\cite{ck3} was proposed in~\cite{starodutsev} as the algebra of symmetries for (2+1) quantum gravity (see also~\cite{Kowconstr}), and compatibility conditions imposed by the Chern--Simons approach to (2+1) gravity have   been recently used~\cite{BHMcqg} in order to identify certain privileged (A)dS quantum deformations~\cite{BHMplb1} (among them, the twisted $\kappa$-(A)dS algebra~\cite{LyakLuki, Dasz1, ktwist, BHMNsigma}).

Concerning (3+1) dimensions, we recall that  in the papers~\cite{Lukierskia, Lukierskib, Giller, Lukierskic} the $\kappa$-Poincar\'e algebra was obtained as a contraction of the Drinfel'd--Jimbo quantum deformation~\cite{Drinfelda, Jimbo} of the $\mathfrak{so}(3,2)$ and $\mathfrak{so}(4,1)$ Lie algebras, by starting from the latter written in the Cartan--Weyl or Cartan--Chevalley basis, and then obtaining suitable real forms of the corresponding quantum complex simple Lie algebras. However, to the best of our knowledge, no explicit expression of the (3+1) $\kappa$-(A)dS algebras in a kinematical basis (rotations $J$, boosts $K$, translations $P$) and including the cosmological constant $\Lambda$ has been presented so far, thus preventing the appropriate physical analysis of the interplay between $\Lambda$ and the quantum deformation. Moreover, explicit expressions for the (3+1) $\kappa$-(A)dS Casimir operators cannot be found in the literature.

The aim of this paper is to fill this gap and to provide the Poisson version of such (3+1) $\kappa$-(A)dS algebra together  with its two deformed Casimirs with non-vanishing cosmological constant: the deformed second-order invariant --which is related to the energy-momentum dispersion relation-- as well as  the deformed fourth-order Casimir, which would be  related to the spin/helicity of the particles~\cite{Bacry, adsJPA}. We recall that in the case $\Lambda=0$ such deformed Pauli--Lubanski  4-vector was obtained for the (3+1) $\kappa$-Poincar\'e algebra in~\cite{ck4}.

From a technical perspective, the quantum (Poisson) algebra will be fully constructed without making use neither of real forms nor of contraction techniques, thus solving the kinematical basis problem since the basis is fully fixed from the initial data. As it was presented in~\cite{dualPL}, by making use of the Poisson version of the quantum duality principle (see~\cite{Drinfelda, STS, Gav, GC} and references therein), it is possible to construct the full Poisson--Hopf algebra structure from the Lie bialgebra that defines the cocomutator $\delta$  ({\em i.e.},~the first-order  of the coproduct $\Delta$)   of the $\kappa$-(A)dS deformation. In short, it can be said that a quantum Poisson algebra is just a Poisson--Lie structure on the dual Lie group $G^\ast$, which has as Lie algebra $g^\ast$ (the one obtained by dualizing the cocommutator $\delta$). In fact, this method was already used in~\cite{dualPL} in order to recover the  Poisson $\kappa$-Poincar\'e algebra in (3+1) dimensions. Moreover, in all the expressions of the  $\kappa$-(A)dS algebra that we will   present, the cosmological constant $\Lambda$ will be introduced as an explicit parameter, and the $\Lambda\to 0$ limit will automatically provide the $\kappa$-Poincar\'e algebra in the so-called bicrossproduct basis~\cite{majid, bicross2, bicross1, bicross3}. 

Evidently, the proper quantum $\kappa$-(A)dS algebra would be obtained by substituting the Poisson brackets here obtained by commutators and by replacing the Poisson algebra generators by noncommuting operators. This would  lead to many ordering ambiguities that have to be solved by the appropriate choice of a symmetrization prescription, which has to be also implemented on the coproduct map (see the discussion in~\cite{dualPL}). Nevertheless, the Poisson approach here presented is enough to get a quite comprehensible description of the deep changes that the non-vanishing cosmological constant generates within the $\kappa$-deformation. In particular, the influence of the cosmological constant on the deformed Casimir operators and on the noncommutative spacetime can be explicitly evaluated. Moreover, the interplay between the cosmological constant and the (1+1) $\kappa$-dS Poisson algebra has been recently presented in~\cite{Giulia1,Giulia2} in the context of  curved momentum spaces and relative locality frameworks~\cite{Kowalski, KowalskiNowak, Freidel, relloc}. Thus, the results here presented provide the tools needed in order to get a deeper insight about the remnants of the quantum geometry that are encoded at the Poisson level in the more realistic (3+1) case. 

The paper is structured as follows. In the next section, the kinematical basis for the (A)dS and Poincar\'e algebras in (3+1) dimensions is revisited  by considering the one-parameter Lie algebra  AdS$_\omega$ whose parameter $\omega$  is related with the cosmological constant in the form $\omega=-\Lambda$. In Section 3, the Poisson $\kappa$-deformation of the (3+1) AdS$_\omega$ algebra is fully constructed as a Poisson--Lie structure on the dual Poisson--Lie group defined by the Lie bialgebra structure that underlies the $\kappa$-defomation. The explicit expresssion of the two Casimir functions is  then obtained and the $\kappa$-Poincar\'e limit is straightforwardly computed. This method can be further applied to the twisted $\kappa$-AdS$_\omega$ algebra arising from a Drinfel'd double structure in (3+1) dimensions, and whose first-order noncommutative spacetime has been recently introduced in~\cite{PLB2015}. A final section including several comments and open problems closes the paper.


\sect{The (3+1) AdS$_\omega$ algebra}

Let us    consider the three  (3+1) Lorentzian Lie algebras as the one-parameter family  AdS$_\omega$, where $\k$ is a real (graded) contraction-deformation  parameter~\cite{adsJPA}. In the   kinematical basis 
$\{P_0,P_a, K_a, J_a\}$  $(a=1,2,3)$ of generators of time translation, space translations, boosts and rotations, respectively, 
the commutation rules for   AdS$_\omega$   read
\be
\begin{array}{lll}
[J_a,J_b]=\epsilon_{abc}J_c ,& \quad [J_a,P_b]=\epsilon_{abc}P_c , &\quad
[J_a,K_b]=\epsilon_{abc}K_c , \\[2pt]
\displaystyle{
  [K_a,P_0]=P_a  } , &\quad\displaystyle{[K_a,P_b]=\delta_{ab} P_0} ,    &\quad\displaystyle{[K_a,K_b]=-\epsilon_{abc} J_c} , 
\\[2pt][P_0,P_a]=\k  K_a , &\quad   [P_a,P_b]=-\k \epsilon_{abc}J_c , &\quad[P_0,J_a]=0  ,
\end{array}
\label{aa}
\ee
where hereafter $a,b,c=1,2,3$ and sum over repeated indices will be    assumed. 
In particular, the contraction parameter $\k$ is related with the cosmological constant in the form $\omega=-\Lambda$, thus AdS$_\omega$ contains the   AdS  Lie algebra $\mathfrak{so}(3,2)$  when  $\k>0$,  the       dS  Lie algebra $\mathfrak{so}(4,1)$  for   $\k<0$, and   the  Poincar\'e one  $\mathfrak{iso}(3,1)$ for  $\k=0$.

The Lie algebra AdS$_\omega$ is endowed with two Casimir operators (see, e.g.,~\cite{adsJPA, casimir}). The quadratic one comes from  the Killing--Cartan form and is given by
\be
{\cal C}
= P_0^2-\>P^2  +\k \left(   \>J^2-\>K^2\right) ,
\label{axa}
\ee
where $P_0^2-\>P^2$
is the square of the energy-momentum
4-vector $(P_0 , \>P )$. For $\k=0$,~\eqref{axa} gives the square of the Poincar\'e invariant rest mass.

The second AdS$_\omega$ Casimir is a fourth-order   invariant that reads
\bea
&&  {\cal W}
=W^2_0-\>{W}^2+\k \left(\>J\producto\>K \right)^2,\nonumber\\[2pt]
&&  W_0= \>J\producto\>P ,
\qquad
 W_a=-  J_a P_0+\epsilon_{abc} K_b P_c ,
\label{axb}
\eea
where  $(W_0,\>{W})$ are just     the components of the 
Poincar\'e Pauli--Lubanski  4-vector. For $\k=0$, the invariant $W^2_0-\>{W}^2$ provides the square of the spin/helicity operator (see~\cite{Bacry, adsJPA}), which in the rest frame it is proportional to the square of the angular momentum. As a consequence, the Casimir ${\cal W}$ (\ref{axb}) generalizes this operator to the case with a non-vanishing cosmological constant.

Notice that  setting $\k=0$ in all the above expressions corresponds  to apply an In\"on\"u--Wigner contraction leading to the flat $\Lambda\to 0$ limit~\cite{LBC}. The Cartan
decomposition of AdS$_\omega$ is given by
$$
 {\rm AdS}_\omega  ={\mathfrak{h}}  \oplus  {\mathfrak{p}} ,\qquad 
{\mathfrak{h}  }={\rm Span}\{ \>K,\>J \}\simeq \mathfrak{so}(3,1) ,\qquad
{\mathfrak{p}  }={\rm Span}\{  P_0,\>P \}  ,
$$
where $\mathfrak{h} $ is the Lorentz subalgebra.
Thus,  the   family   of the  three    (3+1)  Lorentzian  symmetric homogeneous
spaces  with constant   sectional  curvature  $\k$  is defined by  the quotient ${\mathbf
{AdS}}^{3+1}_\k \equiv {\rm  SO}_{\k}(3,2)/{\rm  SO}(3,1) $, where the Lie groups ${\rm  SO}(3,1) $  and ${\rm SO}_{\k}(3,2)$ have  ${\mathfrak{h}} $ and  AdS$_\omega$ as Lie algebras,   respectively. The specific  spaces  are:
\begin{itemize}
\item $\k>0, \Lambda<0$: AdS spacetime ${\mathbf
{AdS}}^{3+1} \equiv  {\rm SO}(3,2)/{\rm  SO}(3,1)$.

\item $\k<0, \Lambda>0$: dS spacetime ${\mathbf
{dS}}^{3+1} \equiv   {\rm  SO}(4,1)/{\rm SO}(3,1)$.

\item $\k=\Lambda=0$: Minkowski spacetime ${\mathbf
M}^{3+1} \equiv  {\rm  ISO}(3,1)/{\rm  SO}(3,1)$.
\end{itemize}
We emphasise that, throughout the paper, the cosmological constant parameter $\omega$ will be explicitly preserved and all the results presented hereafter will hold for any value of $\omega$.


\sect{The Poisson $\kappa$-deformation of the (3+1) AdS$_\omega$ algebra}

The approach presented in~\cite{dualPL} for the construction of the Poisson version of quantum algebras is fully determined by the first-order (in the deformation parameter $z$) of the quantum deformation, which is provided by the cocommutator map $\delta$. In our case,  the classical $r$-matrix for the $\kappa$-deformation of the (3+1) AdS$_\omega$, given in terms of the kinematical generators, is~\cite{LBC}
\be
r=z \left( K_1 \wedge P_1 + K_2 \wedge P_2+ K_3 \wedge P_3 + \sqrt{\k}J_1 \wedge J_2 \right),
\label{ab}
\ee
and from $\delta(X)=[ X \otimes 1+1\otimes X ,  r]$ we obtain the $\kappa$-cocommutator map
\begin{align}
\delta(P_0)&=0,\qquad \delta(J_3)=0 , \nonumber \\
\delta(J_1)&=z \sqrt{\k}   J_1 \wedge J_3  ,\qquad 
\delta(J_2)=z \sqrt{\k}   J_2 \wedge J_3   ,\nonumber \\
\delta(P_1)&=z \left(P_1 \wedge P_0 -\k   J_2  \wedge  K_3+  \k J_3 \wedge K_2 + \sqrt{\k}J_1 \wedge P_3 \right) , \nonumber\\
\delta(P_2)&=z \left(P_2 \wedge P_0 - \k  J_3  \wedge K_1+ \k J_1 \wedge K_3 + \sqrt{\k}  J_2 \wedge P_3 \right) ,\nonumber\\
\delta(P_3)&=z \left(  P_3 \wedge P_0- \k J_1\wedge  K_2  + \k J_2 \wedge K_1  - \sqrt{\k} J_1  \wedge P_1 - \sqrt{\k}  J_2  \wedge P_2 \right), \label{ac}\\
\delta(K_1)&=z \left(K_1 \wedge P_0 + J_2 \wedge P_3 - J_3 \wedge  P_2  +\sqrt{\k}  J_1 \wedge K_3  \right), \nonumber\\
\delta(K_2)&=z \left(   K_2 \wedge P_0+ J_3 \wedge P_1 - J_1  \wedge P_3 +\sqrt{\k} J_2 \wedge K_3   \right) ,\nonumber \\
\delta(K_3)&=z \left(  K_3 \wedge P_0 + J_1 \wedge P_2 - J_2  \wedge P_1  - \sqrt{\k} J_1   \wedge K_1  -\sqrt{\k}  J_2  \wedge  K_2 \right)  .
\nonumber
\end{align}
Here the zero cosmological constant limit corresponds to  setting $\k= 0$, and the quantum deformation parameters $q$, $\kappa$ and $z$  are related  as $q=e^z$ and $  z=1/\kappa$.
 
Now, the quantum duality principle~\cite{Drinfelda, STS, Gav, GC} implies that the Poisson version of the quantum AdS$_\omega$ algebra is just a Poisson--Lie structure on the dual group $G_\k^\ast$, whose Lie algebra $\mathfrak{g}_\k^\ast$ is obtained as the dual of the cocommutator map~\eqref{ac}.  Moreover, the coproduct map for $\kappa$-AdS$_\omega$ is exactly the dual of the product on $G_\k^\ast$, {\em i.e.}, the multiplication law for the dual group  in terms of the corresponding  local coordinates. 
Therefore, if $\{p_0,\>p,\>k,\>j\}$ is the basis for the dual coordinates to $\{P_0,\>P,\>K,\>J\}$, then $\delta$ (\ref{ac})  defines 
the following (solvable) 10-dimensional dual Lie algebra $\mathfrak{g}_\k^\ast$:
\be
\begin{array}{lll}
\left[ j_1,j_2 \right]=0 ,& \quad \left[ j_1,j_3 \right]= z\sqrt{\k}  j_1   , &\quad
 \left[ j_2,j_3 \right]=z\sqrt{\k}  j_2, \\[2pt]
\left[ j_1, p_1 \right]= - z \sqrt{\k}  p_3 ,& \quad \left[j_1,  p_2 \right]= z k_3   , &\quad
\left[j_1,  p_3 \right]= -z (k_2-\sqrt{\k}   p_1)   , \\[2pt]
\left[ j_2, p_1 \right]= -z k_3  ,& \quad \left[ j_2 , p_2\right]=- z\sqrt{\k}   p_3   , &\quad
 \left[ j_2, p_3 \right]=z(k_1 + \sqrt{\k}   p_2)  , \\[2pt]
 \left[ j_3,p_1 \right]= z k_2  ,& \quad \left[ j_3,p_2 \right]= -z k_1     , &\quad
 \left[ j_3, p_3 \right]=0 , \\[2pt]
\left[ j_1 ,k_1 \right]= -z \sqrt{\k}  k_3  ,& \quad  \left[ j_1,k_2 \right]=- z {\k} p_3     , &\quad
 \left[ j_1,k_3 \right]=z (\sqrt{\k} k_1+ {\k} p_2)   , \\[2pt]
\left[ j_2 ,k_1 \right]=  z  {\k}  p_3  ,& \quad  \left[ j_2,k_2 \right]=- z \sqrt{\k} k_3     , &\quad
 \left[ j_2,k_3 \right]=z (\sqrt{\k} k_2- {\k} p_1)   , \\[2pt]
\left[ j_3 ,k_1 \right]=  -z  {\k} p_2  ,& \quad  \left[ j_3,k_2 \right]= z {\k} p_1     , &\quad
 \left[ j_3,k_3 \right]=0 , \\[2pt]
\left[ k_a ,p_0 \right]=  z k_a  ,& \quad  \left[ k_a,k_b \right]= 0     , &\quad
 \left[ k_a,p_b \right]=0 , \\[2pt]
\left[ p_a ,p_0 \right]=  z p_a  ,& \quad  \left[ p_a,p_b \right]= 0     , &\quad
 \left[ p_0,j_a \right]=0  .
 \end{array}
\label{ba}
\ee
Indeed, the limit $z\to 0$ gives rise to an Abelian Lie algebra whose group law is additive for all coordinates, thus giving rise to the undeformed (primitive) coproduct $\Delta(X)=X\otimes 1+1\otimes X $ for the non-deformed quantum algebra.

Therefore, at the Poisson level the deformation induces a non-Abelian dual group $G_\k^\ast$, whose group law provides the explicit coproduct for the Poisson--Hopf algebra AdS$_\omega$. Since
the adjoint representation $\rho$ of $\mathfrak{g}_\k^\ast$ is faithful,  we can use it to
obtain the corresponding 10-dimensional matrix representation of a generic Lie group element in the form:
\begin{align}
G_\k^\ast&=\exp \left(-J_3 \rho(j_3)^t \right) \exp \left(-J_2 \rho(j_2)^t \right) \exp \left(-J_1 \rho(j_1)^t\right) \left(-K_3 \rho(k_3)^t \right)  
\exp \left(-K_2 \rho(k_2)^t \right) \nonumber  \\
&\qquad  \exp \left( -K_1 \rho(k_1)^t \right) \exp \left(-P_3 \rho(p_3)^t \right)
\exp \left(-P_2 \rho(p_2)^t \right) \exp \left( -P_1 \rho(p_1)^t \right) \exp \left(-P_0 \rho(p_0)^t \right) .
\label{bb}\nonumber
\end{align}

A long but straightforward computation (see~\cite{dualPL, BMR} for details abouth the specific procedure) leads to the group law for $G_\k^\ast$, which can be written as the following coproduct for the Poisson AdS$_\omega$ algebra:
\bea
\Delta ( P_0 ) \!\!\!&=&\!\!\!  P_0 \otimes 1+1 \otimes P_0   ,\qquad \Delta  ( J_3  ) =  J_3 \otimes 1 +1 \otimes J_3 , \nonumber\\
\Delta  ( J_1  ) \!\!\!&=&\!\!\!     J_1 \otimes e^{z\sqrt{\k} J_3} +1 \otimes J_1 ,\qquad \Delta ( J_2 ) = J_2 \otimes e^{z\sqrt{\k} J_3}+1 \otimes J_2  , \nonumber
\eea
\bea
\Delta ( P_1 )  \!\!\!&=&\!\!\!  P_1\otimes   \cosh (z\sqrt{\k} J_3) +e^{-z P_0}  \otimes P_1 
- \sqrt{\k} K_2 \otimes     { \sinh (z\sqrt{\k} J_3) }  \nonumber \\
&&- z\sqrt{\k}  P_3  \otimes J_1 + z {\k} K_3 \otimes J_2  + z^2 \k \left(\sqrt{\k} K_1-P_2 \right)  \otimes J_1 J_2 e^{-z\sqrt{\k}  J_3}   \nonumber\\
&&- \frac 12 z^2 {\k}   \left( \sqrt{\k} K_2+P_1   \right)  \otimes      \left( J_1^2-J_2^2 \right)    e^{-z\sqrt{\k} J_3}   , \nonumber \\
\Delta ( P_2 ) \!\!\!&=&\!\!\!  P_2\otimes  \cosh( z\sqrt{\k} J_3) +e^{-z P_0}  \otimes P_2
+ \sqrt{\k} K_1 \otimes    { \sinh (z\sqrt{\k} J_3) }  \nonumber \\
&&- z\sqrt{\k}  P_3  \otimes J_2 - z {\k} K_3 \otimes J_1  - z^2 \k \left(\sqrt{\k} K_2+P_1 \right)  \otimes J_1 J_2 e^{-z\sqrt{\k}  J_3}    \nonumber\\
&& - \frac 12 z^2 {\k}    \left(    \sqrt{\k} K_1- P_2 \right)  \otimes     \left( J_1^2-J_2^2 \right)       e^{-z\sqrt{\k} J_3}  , \nonumber \\
\Delta ( P_3 )  \!\!\!&=&\!\!\!  P_3  \otimes 1 +e^{-z P_0}  \otimes P_3  + z  \left( {\k}  K_2+\sqrt{\k}  P_1 \right)  \otimes J_1 e^{-z \sqrt{\k} J_3}  \nonumber \\
&& -z    \left( {\k}  K_1-\sqrt{\k}  P_2 \right) \otimes  J_2 e^{-z \sqrt{\k}  J_3}  ,  \nonumber 
\eea
\bea
\Delta ( K_1 ) \!\!\!&=&\!\!\!  K_1\otimes     \cosh( z\sqrt{\k} J_3) +e^{-z P_0}  \otimes K_1 +  P_2 \otimes       \frac{ \sinh (z\sqrt{\k} J_3) } {\sqrt{\k}}  \nonumber \\
&&- z P_3  \otimes J_2 - z\sqrt{\k} K_3 \otimes J_1  - z^2  \left(\k K_2+\sqrt{\k}P_1 \right)  \otimes J_1 J_2 e^{-z\sqrt{\k}  J_3}    \nonumber\\
&& - \frac 12 z^2   \left(  {\k}  K_1-  \sqrt{\k}  P_2 \right) \otimes  \left( J_1^2-J_2^2 \right)    e^{-z\sqrt{\k} J_3} ,  \nonumber \\
\Delta ( K_2 ) \!\!\!&=&\!\!\!  K_2\otimes    \cosh (z\sqrt{\k} J_3) +e^{-z P_0}  \otimes K_2 -  P_1 \otimes      \frac{ \sinh (z\sqrt{\k} J_3) } {\sqrt{\k}} \nonumber \\
&&+ z P_3  \otimes J_1 - z\sqrt{\k} K_3 \otimes J_2  - z^2  \left(\k K_1-\sqrt{\k}P_2 \right)  \otimes J_1 J_2 e^{-z\sqrt{\k}  J_3}   \nonumber\\
&&+\frac 12 z^2 \left(\k K_2+\sqrt{\k}  P_1  \right)  \otimes   \left( J_1^2-J_2^2 \right) e^{-z \sqrt{\k}J_3}   ,\nonumber \\
\Delta ( K_3 )  \!\!\!&=&\!\!\!  K_3 \otimes 1+e^{-z P_0}  \otimes K_3 +  z    (\sqrt{\k}  K_1- P_2) \otimes J_1 e^{-z \sqrt{\k}  J_3}  \nonumber\\
&&  + z  (\sqrt{\k}  K_2+P_1) \otimes  J_2 e^{-z \sqrt{\k} J_3}   .
\label{bc}
\eea 
We stress that we have obtained a two-parametric deformation, which is ruled by a ``quantum'' deformation parameter $z=1/\kappa$ together with a ``classical'' deformation parameter $\k$ (the cosmological constant). Notice also that this coproduct is written in a ``bicrossproduct-type'' basis that generalizes the one corresponding to the  (2+1) $\kappa$-AdS$_\omega$ algebra~\cite{BHMNsigma}. 

The Poisson brackets compatible with the previous coproduct as a Poisson--Hopf algebra are given, by Drinfeld's theorem~\cite{DriPL}, as the unique Poisson--Lie structure on $G_\k^\ast$ whose tangent Lie bialgebra is given by the dual of the AdS$_\omega$ commutation rules~\eqref{aa}. Since this dual Lie bialgebra is not a coboundary one, we have no Sklyanin bracket available and the Poisson algebra relations have to be obtained by making use of the computational procedure introduced in~\cite{dualPL}. In particular, we will assume that such Poisson tensor is quadratic in the ``elementary" functions that appear in the previous coproduct, namely
\bea
&&  \left\{ 1, P_0,P_1,P_2,P_3,K_1,K_2,K_3,J_1,J_2,J_3,e^{-z P_0},e^{\pm z  \sqrt{\k} J_3},  J_1 e^{-z  \sqrt{\k}  J_3}, J_2 e^{- z  \sqrt{\k}  J_3}  ,   \right. \nonumber\\
&&\qquad\qquad  \left.  J_1^2 e^{-z  \sqrt{\k} J_3},  J_2^2 e^{- z  \sqrt{\k} J_3} , J_1 J_2 e^{-z   \sqrt{\k}   J_3}\right\} ,\nonumber
\eea
and  the most general antisymmetric quadratic Poisson bracket 
depending on these functions has to be constructed. Afterwards, we impose on this bracket the Poisson homomorphism condition for the deformed $\Delta$~\eqref{bc} that we have previously obtained. This requirement gives rise to a huge system
of linear equations, that can be solved by using a symbolic manipulation program.
Finally, we require that the linearization
of this tentative Poisson tensor gives back the AdS$_\omega$ brackets (\ref{aa}). After enforcing
this last requirement we get the uniquely determined Poisson brackets, which fulfill the Jacobi identity and are explicitly given by: 
$$
\begin{array}{lll}
\multicolumn{3}{l}
{\displaystyle {\left\{ J_1,J_2 \right\}= \frac{e^{2z\sqrt{\k}J_3}-1}{2z\sqrt{\k} } - \frac{z\sqrt{\k}}{2} \left(J_1^2+J_2^2\right)  ,\qquad
\left\{ J_1,J_3 \right\}=-J_2 ,\qquad
\left\{ J_2,J_3 \right\}=J_1 , } } \\[8pt]
\left\{ J_1,P_1 \right\}=z \sqrt{\k} J_1 P_2,  & \quad \left\{ J_1,P_2 \right\}=  P_3-z \sqrt{\k} J_1 P_1   ,  & \quad \left\{J_1, P_3 \right\}=  -P_2  ,  \\[4pt]
 \left\{ J_2,P_1 \right\}=-P_3+z \sqrt{\k} J_2 P_2,  & \quad   \left\{ J_2,P_2 \right\}=  -  z \sqrt{\k} J_2 P_1 ,  & \quad   \left\{ J_2,P_3 \right\}=P_1    ,  \\[4pt]
  \left\{J_3, P_1 \right\}= P_2,  & \quad \left\{ J_3,P_2 \right\}=-P_1 ,  & \quad \left\{ J_3,P_3 \right\}=  0 ,  \\[4pt]
\left\{ J_1,K_1 \right\}=z  \sqrt{\k} J_1 K_2 ,  & \quad \left\{ J_1,K_2 \right\}=  K_3 -z  \sqrt{\k} J_1 K_1   ,  & \quad \left\{J_1, K_3 \right\}=  -K_2  ,  \\[4pt]
 \left\{ J_2,K_1 \right\}= -K_3+z \sqrt{\k}  J_2 K_2,  & \quad   \left\{ J_2,K_2 \right\}= - z  \sqrt{\k} J_2 K_1  ,  & \quad   \left\{ J_2,K_3 \right\}=K_1    ,  \\[4pt]
  \left\{J_3, K_1 \right\}= K_2 ,  & \quad \left\{ J_3,K_2 \right\}=-K_1 ,  & \quad \left\{ J_3,K_3 \right\}=  0 ,  \\[4pt]
 \left\{ K_{a}, P_0 \right\}=P_a   ,  & \quad \left\{ P_0, P_a \right\}=\k K_a ,  & \quad \left\{ P_0,J_a \right\}=  0 ,  \\[4pt]
 \end{array}
\label{be}
$$
\vskip-0.5cm
 \begin{eqnarray*}
 \left\{ K_1,P_1 \right\} \!\!\!&=&\!\!\!   \frac{1}{2z} \left(  \cosh(2z\sqrt{\k} J_3) - e^{-2zP_0}    \right)  +\frac{z^3\k^2}{4} e^{-2z\sqrt{\k}J_3}\left(J_1^2+J_2^2\right)^2+\frac{z}{2} \left( P_2^2+P_3^2-P_1^2\right)  \\
&&
+\frac{z\k}{2}  \left[  K_2^2+K_3^2-K_1^2 +J_1^2 \left(1-e^{-2z \sqrt{\k}  J_3} \right)+ J_2^2 \left(1+e^{-2z \sqrt{\k}  J_3} \right)\right]  ,\\
\left\{ K_2,P_2 \right\}\!\!\!&=&\!\!\! \frac{1}{2z} \left( \cosh(2z \sqrt{\k}  J_3) - e^{-2zP_0} \right) +\frac{z^3\k^2}{4} e^{-2z  \sqrt{\k}   J_3}\left(J_1^2+J_2^2\right)^2+\frac{z}{2} \left(P_1^2+P_3^2 -P_2^2\right)\\
&&
+\frac{z\k}{2} \left[  K_1^2+K_3^2 -K_2^2 +J_1^2 \left(1+e^{-2z \sqrt{\k}  J_3} \right)+ J_2^2 \left(1-e^{-2z \sqrt{\k}  J_3} \right)\right] ,\\
\left\{ K_3,P_3 \right\}\!\!\!&=&\!\!\! \frac{1-e^{-2z P_0}}{2z}  +\frac{z}{2} \left[(P_1+ \sqrt{\k} K_2)^2 +(P_2- \sqrt{\k} K_1)^2-P_3^2  -\k K_3^2   \right]  \\
&&
+z \k e^{-2z  \sqrt{\k}J_3} \left(J_1^2 + J_2^2\right)  ,\\
 \end{eqnarray*}
 \vskip-1.3cm
\begin{eqnarray*}
\left\{ P_1, K_2 \right\} \!\!\!&=&\!\!\! z \left( P_1 P_2+\k K_1 K_2 - \sqrt{\k} P_3 K_3 +\k J_1 J_2 e^{-2 z\sqrt{\k} J_3} \right) , \\
\left\{ P_2, K_1 \right\}\!\!\!&=&\!\!\!  z \left(P_1 P_2 + \k K_1 K_2+  \sqrt{\k}   P_3 K_3+ \k J_1 J_2 e^{-2z \sqrt{\k}  J_3} \right)  ,\\
\left\{ P_1, K_3 \right\}\!\!\!&=&\!\!\!  \frac12  \sqrt{\k} J_1\left(  1- e^{-2z \sqrt{\k} J_3}  \left[1-z^2\k \left(J_1^2+J_2^2 \right) \right]\right)    + z \left(  P_1 P_3 +\k K_1 K_3+ \sqrt{\k}  K_2 P_3\right) ,\\
\left\{ P_3, K_1 \right\}\!\!\!&=&\!\!\! \frac12  \sqrt{\k} J_1\left(  1- e^{-2z \sqrt{\k} J_3}  \left[1-z^2\k \left(J_1^2+J_2^2 \right) \right]\right)    +z \left(P_1 P_3+\k K_1 K_3 - \sqrt{\k}  P_2 K_3 \right) ,\\
\left\{ P_2, K_3 \right\}\!\!\!&=&\!\!\! \frac12  \sqrt{\k} J_2\left(  1- e^{-2z \sqrt{\k} J_3}  \left[1-z^2\k \left(J_1^2+J_2^2 \right) \right]\right) + z \left(P_2 P_3+\k K_2 K_3  - \sqrt{\k} K_1 P_3 \right) , \\
\left\{ P_3, K_2 \right\}\!\!\!&=&\!\!\! \frac12  \sqrt{\k} J_2\left( 1- e^{-2z \sqrt{\k} J_3}  \left[1-z^2\k \left(J_1^2+J_2^2 \right) \right]\right)       + z \left(P_2 P_3+\k K_2 K_3 +   \sqrt{\k}  P_1 K_3 \right) ,\\
\end{eqnarray*}
\vskip-1.3cm
\begin{eqnarray*}
&&\left\{ K_1, K_2 \right\}=-\frac{  \sinh(2 z \sqrt{\k} J_3)}{2z \sqrt{\k} }- \frac{z \sqrt{\k} }2  \left( J_1^2+J_2^2+ 2 K_3^2 \right)- \frac{z^3 \k^{3/2}}4 \, e^{-2z \sqrt{\k}  J_3} \left(J_1^2+J_2^2 \right)^2 , \\
&& \left\{ K_1, K_3 \right\}=\frac12 J_2 \left( 1+e^{-2z  \sqrt{\k}  J_3}\left[  1 +  {z^2 \k}  \left( J_1^2 +J_2^2 \right)  \right]\right)+z  \sqrt{\k}  K_2 K_3 , \\
&& \left\{ K_2, K_3 \right\}=-\frac12 J_1 \left( 1+e^{-2z \sqrt{\k}  J_3}\left[ 1+  {z^2\k }   \left( J_1^2 +J_2^2 \right) \right] \right) -z  \sqrt{\k}  K_1 K_3  ,
\end{eqnarray*}
\vskip-0.75cm
\begin{eqnarray}
&&\left\{ P_1, P_2 \right\}=-\k\, \frac{  \sinh(2 z \sqrt{\k} J_3)}{2z \sqrt{\k} }- \frac{z \sqrt{\k} }2  \left( 2 P_3^2+\k (J_1^2+J_2^2)  \right)- \frac{z^3 \k^{5/2}}4 e^{-2z\sqrt{\k}  J_3} \left(J_1^2+J_2^2 \right)^2  , \nonumber \\
&&\left\{ P_1, P_3 \right\}=\frac12\k J_2 \left( 1+e^{-2z\sqrt{\k}  J_3} \left[1+  {z^2\k}   \left( J_1^2 +J_2^2 \right) \right]  \right) + z \sqrt{\k}  P_2 P_3 ,\nonumber \\
&& \left\{ P_2, P_3 \right\}=-\frac12\k J_1 \left( 1+e^{-2z\sqrt{\k}  J_3} \left[1+  {z^2\k}   \left( J_1^2 +J_2^2 \right) \right]  \right)  -z\sqrt{\k}  P_1 P_3 .\label{ccom}
\end{eqnarray}

This deformed Poisson algebra, together with the deformed coproduct $\Delta$ (\ref{bc}), provides the Poisson version of the $\kappa$-AdS$_\omega$ algebra in (3+1) dimensions. Moreover, the deformed counterpart of the second-order Casimir invariant (\ref{axa}) is found to be
\bea
{\cal C}_z\!\!\!&=&\!\!\! \frac 2{z^2}\left[ \cosh (zP_0)\cosh(z\sqrt{\k} J_3)-1 \right]+\k \cosh(z P_0) (J_1^2+J_2^2) e^{-z \sqrt{\k}  J_3}\nonumber\\
&& -e^{zP_0} \left( \mathbf{P}^2 +\k \mathbf{K}^2 \right)   \left[ \cosh(z\sqrt{\k}  J_3)+ \frac { z^2\k}2  (J_1^2+J_2^2)  e^{-z \sqrt{\k}  J_3} \right]\label{casa}\\
&&+2\k e^{zP_0}  \left[ \frac{\sinh(z \sqrt{\k}J_3)}{\sqrt{\k}}\R_3+ z \left( J_1\R_1 +J_2 \R_2+  \frac {z \sqrt{\k}}2 (J_1^2+J_2^2) \R_3 \right)  e^{-z \sqrt{\k}  J_3} \right],
\nonumber
\eea
where $\R_a=\epsilon_{abc} K_b P_c$.  The deformation of the fourth-order ``Pauli--Lubanski'' invariant  (\ref{axb}) leads to the (quite complicated) expression \begin{eqnarray}
&&\!\!\!\! \!\!\!\! \!\!  {\cal W}_z=-  \frac  {\sinh^2(z P_0) }{z^2}    \left[  \frac {\sinh^2(z  \sqrt{\k} J_3)} {z^2 \omega}+ \frac1 {2 }\, \bigl(1+ e^{-2 z \sqrt{\k}  J_3} \bigr)(J_1^2 + J_2^2) \right]  \nonumber\\
&&\!\!  +  \frac  { \sinh^2(z  \sqrt{\k} J_3)} {2 z^2 \omega}\bigl[  2 e^{2z P_0} (P_3^2+\k K_3^2)  - (e^{2z P_0} -1) (\>P^2 + \k \>K^2)   \bigr] \nonumber\\
&&\!\!  - \frac 14 e^{2z P_0} \left[ \frac{ \sinh(z  \sqrt{\k} J_3)}{ \sqrt{\k} } \, (\>P^2+ \k \>K^2)- 2 R_3 \cosh(z  \sqrt{\k} J_3)  \right]^2
\nonumber \\
&&\!\! +\frac{ (e^{2z P_0} -1)  \sinh(2 z  \sqrt{\k} J_3)}{2 z^2 \sqrt{\k}}\, R_3
  -  e^{2z P_0} \bigl[   R_1^2+ R_2^2 - 2  e^{-2 z \sqrt{\k}  J_3} J_1 J_2 (P_1 P_2+\k K_1 K_2)      \bigr] 
\nonumber\\
&&\!\! +  e^{2z P_0} \left[  \frac {1- e^{-2 z \sqrt{\k}  J_3}  }{z \sqrt{\k} } + z \sqrt{\k}\,  e^{-2 z \sqrt{\k}  J_3}  (J_1^2 + J_2^2)  \right]  \!   \bigl[ (J_1P_1  + J_2P_2) P_3 +\k( J_1K_1 +J_2K_2)K_3\bigr]\nonumber\\
&&\!\!  + \frac{e^{2z P_0} -1}{2 z} \bigl(1+ e^{-2 z \sqrt{\k}  J_3} \bigr) (J_1 R_1 + J_2 R_2)+ e^{2z P_0} e^{-2 z \sqrt{\k}  J_3} \left[ J_1^2   (P_1^2 + \k K_1^2  )+J_2^2 \left(P_2^2 + \k K_2^2 \right )  \right]  
\nonumber\\
&&\!\!  -\frac 14 (J_1^2+ J_2^2) \left[  (e^{2z P_0}-1)  ( \LL_-  +  e^{-2 z \sqrt{\k}  J_3} (\>P^2+\k \>K^2)  ) + 2  e^{2z P_0}(e^{-2 z \sqrt{\k}  J_3}-1)(P_3^2+\k K_3^2)\right] \nonumber \\
&&\!\!  + \frac z 2\,  e^{2z P_0}  e^{-2 z \sqrt{\k}  J_3} (J_1R_1+J_2 R_2) \bigl[ \LL_+ +  e^{2 z \sqrt{\k}  J_3} \LL_-  +\k (1 -e^{-2z P_0} )(J_1^2+J_2^2)  \bigr] \nonumber \\
&&\!\! - z^2 \k \, e^{2z P_0}e^{-2 z \sqrt{\k}  J_3} (J_1R_1+J_2R_2)^2 - \frac{z^2}8 e^{2z P_0} (J_1^2+J_2^2)\LL_- (\LL_- +e^{-2 z \sqrt{\k}  J_3}  \LL_+)\nonumber \\
&&\!\! - \frac{z^2}4 \k\, e^{-2 z \sqrt{\k}  J_3} (J_1^2+ J_2^2)^2 \left[ \frac  {\sinh^2(z P_0) }{z^2}  -   e^{2z P_0} (P_3^2+\k K_3^2) +\frac{ e^{2z P_0} -1}{2}\, \LL_-\right]\nonumber \\
&&\!\!  +\frac{z^3}2\k\, e^{2z P_0} e^{-2 z \sqrt{\k}  J_3}(J_1^2+ J_2^2) \LL_-\left[  J_1R_1+ J_2 R_2 - \frac{z}{8}(J_1^2+ J_2^2) \LL_- \right]
, 
 \label{casb}
\end{eqnarray}
where  we have used the notation $\LL_\pm =\>P^2+\k \>K^2\pm 2  \sqrt{\k} R_3 $. Both the ``classical'' and ``flat''  limits $z\to 0$ and $\k\to 0$ are always well--defined.

Indeed, what we have obtained is a commutative Poisson--Hopf algebra, whose quantization has to be performed in order to obtain the proper $\kappa$-AdS$_\omega$ quantum algebra. Once this had been completed, a (presumably nonlinear and complicated) change of basis should exist between the kinematical basis and the Cartan--Weyl  or Cartan--Chevalley basis used in~\cite{Lukierskia, Lukierskib, Giller, Lukierskic}. Nevertheless, most of the physically relevant features of this quantum deformation can already be  extracted from the kinematical Poisson structure here obtained.


\subsect{The Poisson $\kappa$-Poincar\'e algebra}

 Under  the $\omega\to 0$ limit  the classical $r$-matrix (\ref{ab}) and the cocommutators (\ref{ac}) reduce to
\bea
&& r=z \left( K_1 \wedge P_1 + K_2 \wedge P_2+ K_3 \wedge P_3   \right) , \nonumber\\
&&\delta(P_0)=0,\qquad \delta(J_a)=0 ,  \qquad \delta(P_a)=z P_a \wedge P_0 , \nonumber \\
&&\delta(K_a)=z \left(K_a \wedge P_0 +  \epsilon_{abc}   P_b \otimes J_c  \right) ,
\label{acc}\nonumber
\eea
which means that the dual group is ``more Abelian'' than the $\omega\neq 0$ one. Therefore the dual group law is simpler, and the $\omega\to 0$ limit of (\ref{bc}) and (\ref{ccom}) lead to the following  coproduct and   Poisson brackets  
\begin{eqnarray}
&& \Delta ( P_0 ) = P_0 \otimes 1+1 \otimes P_0 ,\nonumber\\
&&  \Delta ( J_a ) =   J_a \otimes 1+1 \otimes J_a , \nonumber\\
&& \Delta ( P_a ) =  P_a \otimes 1+e^{-z P_0}  \otimes P_a  ,\nonumber \\
&& \Delta ( K_a ) =  K_a \otimes  1+e^{-z P_0}  \otimes K_a+z \epsilon_{abc}   P_b \otimes J_c  ,
\nonumber
\end{eqnarray}
\be
\begin{array}{lll}
\left\{ J_a,J_b\right\}=\epsilon_{abc}J_c ,& \quad \left\{  J_a,P_b\right\}=\epsilon_{abc}P_c , &\quad
\left\{  J_a,K_b\right\}=\epsilon_{abc}K_c , \\[2pt]
\displaystyle{
  \left\{ K_a,P_0\right\}=P_a  } , &\quad\displaystyle{  \left\{ K_a,K_b\right\}=-\epsilon_{abc} J_c    } ,    &\quad\displaystyle{   \left\{ P_0,J_a\right\}=0  } , 
\\[2pt]
\left\{ P_0,P_a\right\}= 0 , &\quad   \left\{ P_a,P_b\right\}=0 , &\quad   \\[2pt]
\multicolumn{3}{l}
{\displaystyle      {   \left\{ K_a, P_b \right\} = \delta_{ab} \left( \frac{1}{2z} \left(1-e^{-2z P_0} \right) +\frac{z}{2} \>P^2 \right)- z P_a P_b  } \, .} 
\end{array}
\label{aa2}\nonumber
\ee
Since all the coordinate  functions Poisson-commute, ordering ambiguities do not exist and the quantization of this Poisson--Hopf algebra is immediate, thus giving exactly the $\kappa$-Poincar\'e algebra in the bicrossproduct basis~\cite{bicross2,  bicross1, bicross3}. The 
deformed  quadratic Casimir function providing the deformed mass-shell condition  is obtained through the $\omega\to 0$ limit of (\ref{casa}); namely
\be
 {\cal C}_z=\frac{2}{z^2}\left[  \cosh(zP_0)-1\right]- e^{z P_0}  \>P^2=\frac{4}{z^2}\sinh^2(zP_0/2) - e^{z P_0}\>P^2.
\label{casdeformeda}
\ee
The    deformed Pauli--Lubanski invariant, coming from the $\omega\to 0$ limit of (\ref{casb}),  can be written in the form
\bea
&&\!\!\!\! \!\!\!\! 
 {\cal W}_z
=\left(  \cosh(zP_0)-  \frac {z^2}4 \, e^{z P_0}   \>P^2\right)W^2_{z,0}-\>{W}_z^2  \,  ,\nonumber\\ 
&&\!\!\!\! \!\!\!\! 
  W_{z,0}=e^{\frac z 2 P_0}\,  \>J\producto\>P ,
\qquad
 W_{z,a}=-  J_a \, \frac{\sinh(z P_0)}{z}+e^{z P_0}\epsilon_{abc} \left( K_b + \frac z 2 \, \epsilon_{bkl}J_k P_l \right)P_c   \, ,
\label{casdeformedb}
\eea
to be compared with (\ref{axb}) with $\k=0$.


\subsect{A twisted $\kappa$-AdS$_\omega$ algebra}

If we add to the $r$-matrix (\ref{ab}) a Reshetikhin twist of the type 
$
r_{\vt}=\vt J_3 \wedge P_0,
$
we are led to consider the two-parametric Lie bialgebra structure generated by
\be
r_{z,\vt}=z \left( K_1 \wedge P_1 + K_2 \wedge P_2+ K_3 \wedge P_3 + \sqrt{\k}J_1 \wedge J_2 \right)+\vt  J_3 \wedge P_0,
\label{da}
\ee
which is just the $r$-matrix that generates a Drinfel'd double quantum deformation of the AdS$_\omega$ algebra in (3+1) dimensions that has been recently considered in~\cite{PLB2015}.  The corresponding cocommutator map is the sum of two terms $\delta_{z,\vt}=\delta+\delta_\vt$ where $\delta$ is given in (\ref{ac}) while the twisted component  $\delta_\vt$ reads
\begin{align}
\delta_\vt(P_0)&=0,\qquad \delta_\vt(J_3)=0 , \nonumber \\
\delta_\vt(J_1)&= -\vt   J_2 \wedge P_0  , \qquad 
\delta_\vt(J_2)=\vt   J_1 \wedge P_0  ,  \nonumber\\
\delta_\vt(P_1)&=-\vt \left( P_2 \wedge P_0 +\k   J_3  \wedge  K_1 \right) , \nonumber\\
\delta_\vt(P_2)&=\vt\left( P_1 \wedge P_0 -\k   J_3  \wedge  K_2 \right) ,\nonumber\\
\delta_\vt(P_3)&=-\vt \k    J_3  \wedge K_3, \nonumber\\
\delta_\vt(K_1)&=-\vt  \left( K_2 \wedge P_0  -   J_3  \wedge  P_1 \right)   , \nonumber\\
\delta_\vt(K_2)&=\vt \left( K_1 \wedge P_0 +  J_3  \wedge  P_2 \right) ,\nonumber \\
\delta_\vt(K_3)&=\vt     J_3  \wedge  P_3   .
\label{accc}
\end{align}

The Poisson version of the corresponding twisted $\kappa$-AdS$_\omega$ algebra can be obtained by following the same method as in the untwisted case, and will be presented elsewhere. In fact, the additional twist contribution~\eqref{accc} to the cocommutator would lead to a dual Poisson--Lie group $G_{\omega,\vt}^\ast$ which would be ``less Abelian" than $G_{\omega}^\ast$, thus leading to a more complicated dual group law. Nevertheless, the Poisson--Lie brackets compatible with such two--parametric  coproduct  will be the same as in the non-twisted case. Again, the Lie bialgebra underlying the  corresponding  Poisson twisted $\kappa$-Poincar\'e algebra is obtained from (\ref{da}) in the vanishing cosmological constant limit $\omega\to 0$ and, its full Hopf structure has been worked out in~\cite{LyakLuki, Dasz1, ktwist, dualPL}.


\sect{Concluding remarks}

Several features of the $\kappa$-(A)dS Poisson--Hopf algebra presented in the previous section are worth to be commented. 
Firstly, it becomes apparent that the existence of a  non-vanishing cosmological constant implies a significantly more complicated Hopf algebra structure.
This will have a deep impact as far as the quantization of the Hopf algebra is concerned, since many ordering ambiguities (which are absent in the $\kappa$-Poincar\'e case) appear. As it was already pointed out in~\cite{dualPL}, such ordering issues can be minimized if we consider a basis in which the deformed coproduct is invariant under the composition of the flip operator $\sigma(X\otimes Y)=Y\otimes X$ with the change $z\to -z$ in the deformation parameter. This ``symmetrical" basis for the $\kappa$-(A)dS$_\omega$ algebra can be found and turns out to be
\bea
&& \tilde{P_0}= P_0,\qquad  \tilde{J_3}=J_3, \qquad \tilde{J_1}=J_1 e^{-\frac{z}{2}  \sqrt{\k} J_3} ,\qquad \tilde{J_2}=J_2 e^{-\frac{z}{2} \sqrt{\k}  }J_3  ,\nonumber \\
&& \tilde{P_1}= \cosh \left( \frac{z}{2}\sqrt{\k} J_3 \right) e^{\frac{z}{2} P_0} P_1 +\sqrt{\k}  \sinh \left( \frac{z}{2}\sqrt{\k} J_3 \right)  e^{\frac{z}{2} P_0} K_2 \nonumber\\
&&\qquad 
+ \frac{z}{2}\sqrt{ \k}
 e^{-\frac{z}{2} (\sqrt{\k} J_3- P_0)} \left(J_1 P_3 - \sqrt{\k} J_2 K_3 \right) ,\nonumber \\
&& \tilde{P_2}= \cosh \left( \frac{z}{2}\sqrt{\k} J_3 \right) e^{\frac{z}{2} P_0} P_2 - \sqrt{\k}  \sinh \left( \frac{z}{2}\sqrt{\k} J_3 \right)  e^{\frac{z}{2} P_0} K_1 \nonumber\\
&&\qquad 
+ \frac{z}{2}\sqrt{ \k} e^{-\frac{z}{2} (\sqrt{\k} J_3- P_0)} \left(J_2 P_3 + \sqrt{\k} J_1 K_3 \right) , \nonumber \\
&&\tilde{P_3}=\left[ P_3 + \frac{z}{2}  \sqrt{\k} \left( J_2 ( \sqrt{\k} K_1-P_2)- J_1 ( \sqrt{\k} K_2+P_1) \right)  \right] e^{\frac{z}{2} P_0} , \nonumber\\
&& \tilde{K_1} =\cosh \left( \frac{z}{2}\sqrt{\k} J_3 \right) e^{\frac{z}{2} P_0}  K_1 - \frac{  \sinh \left( \frac{z}{2}\sqrt{\k} J_3 \right) }{\sqrt{\k}} e^{\frac{z}{2} P_0}   P_2 \nonumber\\
&&\qquad 
 + \frac{z}{2} e^{-\frac{z}{2} (\sqrt{\k} J_3- P_0)} \left(  J_2 P_3 + \sqrt{\k}   J_1 K_3 \right) , \nonumber\\
  && \tilde{K_2} =\cosh \left( \frac{z}{2}\sqrt{\k} J_3 \right) e^{\frac{z}{2} P_0}  K_2 + \frac{  \sinh \left( \frac{z}{2}\sqrt{\k} J_3 \right) }{\sqrt{\k}} e^{\frac{z}{2} P_0}   P_1 \nonumber\\
&&\qquad 
 - \frac{z}{2} e^{-\frac{z}{2} (\sqrt{\k} J_3- P_0)} \left(  J_1 P_3 - \sqrt{\k}   J_2 K_3 \right) , \nonumber\\
&&\tilde{K_3}=\left[ K_3- \frac{z}{2} \left( J_1 ( \sqrt{\k}  K_1-P_2)+ J_2 (\sqrt{\k}   K_2+P_1) \right)  \right] e^{\frac{z}{2} P_0} .
\nonumber
\eea
Under this change of basis functions we have, for instance, that
\be
\Delta(\tilde{J_1})= \tilde{J_1} \otimes e^{\frac{z}{2} \tilde{J_3}} + e^{-\frac{z}{2} \tilde{J_3}} \otimes \tilde{J_1},
\qquad 
\Delta(\tilde{J_2})= \tilde{J_2} \otimes e^{\frac{z}{2} \tilde{J_3}} + e^{-\frac{z}{2} \tilde{J_3}} \otimes \tilde{J_2}.
\label{copsym}
\ee
All the remaining new coproducts in this basis can  straightforwardly be written, but we omit them for the sake of brevity. The full solution of the quantization problem is currently under investigation.

It is also worth mentioning that expressions~\eqref{copsym} reflect that  the rotation sector of the (3+1) $\kappa$-(A)dS algebra is a quantum $\mathfrak{so}(3)$ subalgebra, which is generated by the $\sqrt{\k}J_1 \wedge J_2$ term in the classical $r$-matrix~\eqref{ab}. This constitutes an essential difference with respect to the Poincar\'e case, since in the limit $\omega\to 0$ this rotation subalgebra becomes a non-deformed one. The same happens for $\omega\neq 0$ in the  (2+1) $\kappa$-(A)dS deformation, which is generated by the classical $r$-matrix
\be
r=z \left( K_1 \wedge P_1 + K_2 \wedge P_2 \right),
\label{ab2+1}\nonumber
\ee
where the cosmological constant does not appear. Moreover, both in the (2+1) and (3+1) cases the Lorentz sector is not a Hopf subalgebra for any value of $\omega$.

On the other hand, it can also be  appreciated that the Casimirs~\eqref{casa} and~\eqref{casb} are profoundly modified by the cosmological constant when they are compared with their $\kappa$-Poincar\'e counterparts~\eqref{casdeformeda} and~\eqref{casdeformedb}. In particular, the quadratic one~\eqref{casdeformeda} giving rise to the $\kappa$-Poincar\'e dispersion relation  is now transformed into~\eqref{casa}, where both boost and rotation generators do contribute. The physical significance of this fact deserves further study, that could be approached by taking into account the results obtained in~\cite{Giulia1,Giulia2} for the (1+1)   $\kappa$-dS case, and the common features with respect to the (2+1) $\kappa$-(A)dS Casimirs that were commented in~\cite{BHMcqg}.

Another important aspect to be considered is the noncommutative $\kappa$-(A)dS spacetime associated with the deformation here presented. This noncommutative spacetime will be a nonlinear algebra whose first-order  can be directly obtained as a Lie subalgebra of the dual Lie algebra~\eqref{ba}. If we identify
$\hat x^0\equiv p_0$ and $\hat x^a\equiv p_a$, we find that the   first-order $\kappa$-(A)dS$_\omega$ noncommutative spacetime reads 
\be
[ \hat x^a, \hat x^0]=z\, \hat x^a       , \qquad
 [\hat x^a, \hat x^b]=0  ,\qquad z=1/\kappa, \qquad a,b=1,2,3.
  \label{cf}
\ee
This is just the well known (3+1) $\kappa$-Minkowski spacetime (see~\cite{kMinkowski, bicross2, kZakr, LukR}), and the cosmological constant does not appear at this first-order level. Higher-order contributions will indeed contain $\omega$, and they will arise when the full Hopf algebra duality for the $\kappa$-(A)dS coproduct~\eqref{bc} is computed (see~\cite{RossanoPLB}). Alternatively, such all--orders (3+1)   noncommutative spacetime can be obtained as a Poisson subalgebra within the Poisson--Lie structure defined by the Sklyanin bracket coming from the classical $r$-matrix~\eqref{ab} (see~\cite{BHMNsigma} for the (2+1) case). In the same manner, the first-order twisted version of this noncommutative spacetime is obtained by adding to~\eqref{cf} the contributions coming from~\eqref{accc}, and yields
\begin{align}
[\hat x^1,\hat x^0]&=z\,\hat x^1+\vt \, \hat x^2 ,  \qquad  [\hat x^2,\hat x^0]=z \,\hat x^2 - \vt\,\hat x^1, \qquad  [\hat x^3,\hat x^0]=z\,\hat x^3, \nonumber \\
\label{nonAdS}
 [\hat x^a,\hat x^b]&=0 ,\qquad  a,b = 1,2,3\, ,
 \nonumber
\end{align}
which is not isomorphic to the (3+1) $\kappa$-Minkowski spacetime (see~\cite{PLB2015}). Again, the cosmological constant will appear for higher-orders within the full commutation rules. We recall that in the (2+1) case the twist guarantees the compatibility with the Chern--Simons approach to gravity~\cite{BHMcqg}. Work on all these problems is in progress and will be presented elsewhere.


\section*{Acknowledgements}

This work was partially supported by Ministerio de Econom\'{i}a y Competitividad (MINECO, Spain) under grants MTM2013-43820-P, and by Junta de Castilla y Le\'on (Spain) under grants BU278U14 and VA057U16. P.N. acknowledges a postdoctoral grant by Junta de Castilla y Le\'on.


\end{document}